\documentstyle[preprint,aps]{revtex}
\begin{document}
\title{ Novel nonreciprocal acoustic effects in antiferromagnets}

\author{ Roser Valent\'{\i}, Claudius Gros} 
\address{   Institut f\"ur
 Physik, Universit\"at Dortmund, 44221 Dortmund, Germany. \\}
\author{V.N. Muthukumar} 
\address{Department of Physics, Brookhaven National Laboratory, 
         Upton, New York 11973-5000 \\}

\date{\today}

\maketitle

\begin{abstract}

  The possible occurrence of nonreciprocal acoustic effects in
 antiferromagnets in the absence of an external
 magnetic field is investigated using both (i)
 a microscopic formulation of the magnetoelastic interaction
 between spins and phonons and (ii) symmetry arguments.  We
 predict for certain antiferromagnets the existence of 
 two new nonreciprocal (non-time invariant) effects:
 A boundary-condition induced nonreciprocal effect
 and the occurrence of transversal phonon modes
 propagating in opposite directions having different
 velocities.  Estimates are given and possible
 materials for these effects to
 be observed are suggested.

\end{abstract}


\paragraph*{Introduction}

  It is well known in optics that the coupling of light to the 
 order parameter in a ferromagnet mediates nonreciprocal
 effects
  such as the Faraday effect \cite {Argyres}.  
The Faraday
 effect in ferromagnets appears as a rotation of the plane of
 polarization of light which is incident along the axis of magnetization.
  An analogous effect in acoustics (the so-called
 {\it {acoustic}} Faraday effect)  was first predicted  
by Kittel \cite {Kittel}. Subsequently, it was verified experimentally 
\cite{Matthews} that in a ferromagnet, circularly polarized
 phonons of different sense have different velocities
 when the direction of propagation is parallel to the axis of magnetization.
Such an effect can be understood in terms of the magnetoelastic 
 coupling between spins and phonons.  Boiteux {\it {et al.}}
\cite{Boiteux} performed in 1971 a detailed study of the
 magnetoacoustic effects in 
  antiferromagnetic Cr$_2$O$_3$.  They found that
 the crystal symmetry of Cr$_2$O$_3$  allows an acoustic Faraday
 effect only in the presence of an external magnetic field.

   The nonreciprocal effects described above rely on 
the existence of a well defined magnetic moment induced either
 by the internal moments in the
case of ferromagnets, or by the application of an
 external magnetic field in the case
 of Cr$_2$O$_3$ \cite{Boiteux}. 
It is not apparent that such effects can occur in ordered
antiferromagnets in the absence of an
 external magnetic field where the total magnetic moment is zero.  
In the case of optics, symmetry arguments can be used to deduce the 
presence of nonreciprocal
 bulk effects \cite{Dzya} as well as a range of 
 characteristic surface effects \cite{Daehn} in antiferromagnets.
Indeed, nonreciprocal effects in antiferromagnets were demonstrated 
recently using the technique of Second Harmonic Generation
 \cite{Fiebig}.   
 A microscopic explanation of these experiments 
 which relies on the symmetry of the crystal and 
 the evaluation of the interactions at the atomic level 
has also been proposed \cite{Muthu,Tanabe}.

 In this Letter we examine 
if it is possible to observe nonreciprocal 
 phonons in antiferromagnets in the absence 
 of an external magnetic field. We find that the answer is
 affirmative but these effects
 are very different from those observed in optics.
 Our calculations show that there is a new boundary-condition induced 
 elliptical-polarization effect in antiferromagnets. In addition
 we predict,  for certain antiferromagnets, the occurrence of
 transversal phonon modes
 propagating in opposite directions having different
 velocities, 
 an effect which
 cannot occur in ferromagnets.

\paragraph*{Ferromagnets}

Kittel \cite{Kittel} was the first to predict the magnetoacoustic
Faraday effect in ferromagnets within a semiclassical 
continuum formulation. For the sake of completeness, 
we present a brief derivation of
Kittel's result within a linear chain model for a ferromagnet.
We take the $\hat z$-axis to be the direction of the
chain of ions corresponding to the main crystallographic axis,
which we assume to have C$_4$ or higher rotational symmetry. The
magnetization is parallel to the $\hat z$-axis.  

We define $u_{n,\gamma}$ and $p_{n,\gamma}$  ($n$ denotes the
site index and $\gamma= x,y$) to be the
canonical displacement and momentum operators 
obeying the commutation relations
$\ [u_{n,\gamma},p_{n^\prime,\gamma^\prime}]=
i\hbar\delta_{n,n^\prime}\delta_{\gamma,\gamma^\prime}$.
The Hamiltonian $H=T+V_{el}+H_{Heis}+H_{sp}$ contains the
unperturbed phonon part
\begin{equation}
T\ =\ {1\over 2m}\sum_{n,\gamma} p_{n,\gamma}^2,\qquad
V_{el}\ =\  {K\over2}\sum_{n,\gamma} \left(u_{n+1,\gamma}-u_{n,\gamma}\right)^2
\label{elastic}
\end{equation}
where $K$ is the elastic constant, and $m$ the mass of the ion.
$H_{Heis}$ corresponds to the Heisenberg Hamiltonian and describes
the unperturbed spin-spin interaction,
\begin{equation}
H_{Heis}\ =\ J \sum_{n} \vec {S}_n \cdot \vec { S}_{n+1}
\label{heisenberg}
\end{equation}

$J<0$ is the ferromagnetic exchange constant and
$ \vec {S}_n$ corresponds to the
spin operator on site $n$, with
$[S_n^\alpha,S_m^\beta]=i\hbar\delta_{n,m}\epsilon_{\alpha\beta\gamma}
S_n^\gamma$ ($\alpha,\beta,\gamma=x,y,z$).
$H_{sp}$ describes
 the magnetoelastic coupling between phonons and spins:
\begin{equation}
H_{sp}\ =\  g\sum_{n,\gamma} \left(u_{n+1,\gamma}-u_{n,\gamma}\right)
              \left(S_{n+1}^\gamma S_{n}^z+S_{n+1}^z S_{n}^\gamma\right).
\label{Hmag}
\end{equation}
  $g$ is the magnetoelastic coupling constant.

 Eq.\ (\ref{Hmag}) is rotational invariant 
around the $\hat{z}$-axis.  A 90-degree
 rotation around the $\hat{z}$-axis would correspond to
 ($u_{n,x} \rightarrow -u_{n,y}$, $u_{n,y} \rightarrow u_{n,x}$,
 $S_n^x \rightarrow -S_n^y$, $S_n^y \rightarrow S_n^x$).
Eq.\ (\ref{Hmag}) is the lattice version of the usual
coupling between components of the strain tensor and
the magnetization \cite{Kittel}. 
This interaction term can also be understood as
 a higher order expansion of the exchange term and
 has its origin in the spin-orbit coupling\cite{Moriya}.
 In order to obtain the time dependence of the variables
$p_{n,\gamma}$, $u_{n,\gamma}$ and
$S_n^{\gamma}$  we consider
the Heisenberg equation of motion $\ i\hbar {d\over dt} A= [A,H]$.
  For simplicity, we work with circular-polarized coordinates
$u_n^{\pm} = u_{n,x} \pm i u_{n,y}$,  $S_n^{\pm} = S_{n}^{x} \pm i
 S_{n}^{y}$   and we assume
a plane wave ansatz
$u_{n}^{\pm}(t)= \exp[i(\omega t-nk)]u^{\pm}$,
$S_n^{\pm}(t)=\exp[i(\omega t -nk)]S^{\pm}$ (the unit cell constant
 is taken to be one) in order to solve the
 equations of motion. We set the magnetization to be along the
$\hat z$-axis
 ($\langle S_n^x\rangle =0$,
$\langle S_n^y\rangle =0$,
$\langle S_n^z\rangle =\langle S^z \rangle$)
and we obtain after eliminating $\dot p_{n,\gamma}$:
\begin{eqnarray}
\omega S^{\pm}=\pm g \langle S^z \rangle ^2 (2 i \sin k) u^{\pm}
&\pm&J \langle S^z \rangle 2 (1 - \cos k)S^{\pm}\label{us1}\\
(-m\omega^2 + 2 K (1- \cos k)) u^{\pm}&=&g \langle S^z \rangle
  (-2 i \sin k) S^{\pm}~.
\label{us2}
\end{eqnarray}

 Combining Eq.\ (\ref{us1}) and (\ref{us2}) we get:
\begin{eqnarray}
\left\{
[-m\omega^2 + 2K(1 - \cos k)][\omega \mp J \langle S^z \rangle
 2(1 - \cos k)] \mp 4g^2 \langle S^z \rangle ^3 (\sin k)^2
\right\}u^{\pm}=0
\label{fdispersion}
\end{eqnarray}
which describes the dispersion relation of the  phonon
 system interacting with the spin system.
 Note that for  $u^+$ and $u^-$   the dispersion relation is
 different implying that for a given frequency $\omega$, right (+)
 and left (-) circular polarized phonons have different wave
 vectors $k^+$ and $k^-$ respectively.

An incoming linear polarized wave
with frequency $\omega$ is decomposed into a linear
superposition of right and left circular polarized waves
with the same frequency $\omega$ but different
wave vectors $k^+/k^-$ via
$\ \exp[i(\omega t-k^+ n)]\,{\bf u}^{+}
+\exp[i(\omega t-k^- n)]\,{\bf u}^{-}$. The physical
amplitudes, determined by the real part of $u$ are given by
\begin{eqnarray}
u_{n,x} &=& u\cos(\omega t-n(k^++k^-)/2)\cos(n(k^--k^+)/2)
\nonumber\\
u_{n,y} &=& u\cos(\omega t-n(k^++k^-)/2)\sin(n(k^--k^+)/2).
\nonumber
\end{eqnarray}
The plane of polarization rotates around the $\hat z$-axis 
\cite{Kittel} and it is easy to obtain from the solution of
Eq.\  (\ref{fdispersion}) the rotatory power
$\ (k^+-k^-)/(k^++k^-)\approx g^2\langle S^z\rangle ^3/(K\omega)$.
  The nonreciprocal
 behavior is a direct consequence of the coupling
 between phonons and spins in a ferromagnet.
 For a typical ferrite the rotatory
 power should be of the order of 1/4 for sound
 waves in the microwave regime ($w \approx 10^9 s^-1$), as was
 corroborated experimentally \cite{Matthews}.


\paragraph*{Antiferromagnets}

As a model for an antiferromagnetic material, we
consider a linear chain with two magnetic
ions ($A/B$) per unit-cell, with $\vec u_{A,n}$
and $\vec u_{B,n}$ being their respective
displacement vectors. 
  A straightforward generalization
of (\ref{elastic}), (\ref{heisenberg})  and (\ref{Hmag}) leads to 

\begin{eqnarray}
T = \frac{1}{2m} \sum_{n,\gamma} \left( p_{A,n, \gamma}^2 +
 p_{B,n, \gamma}^2\right) \nonumber \\
V_{el} = \frac{K_1}{2} \sum_{n, \gamma}
(u_{B,n,\gamma} - u_{A,n,\gamma})^2
 + \frac{K_2}{2} \sum_{n,\gamma} (u_{A,n+1,\gamma} - u_{B,n,\gamma})^2~.
\end{eqnarray}
for the phonon part.
$K_1$ and $K_2$ are, respectively,
 the intra- and intercell elastic constants.
\begin{eqnarray}
H_{Heis} =  \sum_{n}\, [J_1 \vec{S}_{A,n}\cdot\vec{S}_{B,n}
 + J_2\vec{S}_{B,n}\cdot\vec{S}_{A,n+1} ]~,
\end{eqnarray}
 corresponds to the Heisenberg term 
with $J_1>0$ and $J_2>0$ (antiferromagnetic).  And
the magnetoelastic coupling is given by
\begin{eqnarray}
H_{sp}^{(g)} = g_1 \sum_{n, \gamma} (u_{B,n,\gamma}-u_{A,n,\gamma})
( S_{B,n}^{\gamma} S_{A,n}^z + S_{B,n}^z S_{A,n}^{\gamma}) \nonumber \\
+ g_2 \sum_{n, \gamma} (u_{A,n+1,\gamma} - u_{B,n,\gamma})
 (S_{A, n+1}^{\gamma} S_{B,n}^z + S_{A, n+1}^z S_{B, n}^{\gamma})
\label{H_sp2}
\end{eqnarray}
where $g_1$ and $g_2$ are, respectively, intra- and intercell
 magnetoelastic coupling constants.
In order to obtain
 the dispersion relation of the phonons interacting with the spin
 system, we set
  the staggered magnetization to be parallel to the $\hat{z}$ axis
  ($\langle S_{A,n}^z\rangle \rightarrow \langle S_A^z \rangle $
 and
 $\langle S_{B,n}^z\rangle \rightarrow - \langle S_A^z \rangle $)
 and solve the equations of motion for
 $\vec{u}_{A/B,n}$,  $\vec{p}_{A/B,n}$
 and $\vec{S}_{A/B,n}$.  

We consider the plane-wave ansatz
${\bf u}_n=\exp[i(\omega t-kn)]\,{\bf u}$ and
 define the vector 
$ {\bf u}_n=(u_{A,n,x},u_{B,n,x};u_{A,n,y},u_{B,n,y}) $.
Eliminating $\dot p_{A/B,n,\gamma}$ and $\dot S_{A/B,n,\gamma}$
we obtain $M {\bf u} = 0$, where the $ 4 \times 4$
  hermitian matrix $M$ is given by:
\[
\left(
\begin{array}{cccc}
 L(\omega)+ 2J s^4 R_k &
-T_k (\omega)
- J s^4 P_k& 2 i s^3 Q_k \omega & 0 \\
-T_{-k}(\omega)
- J s^4 P_{-k} &
 L(\omega)+ 2J s^4 R_k & 0 &
-2i s^3 Q_k \omega \\
-2i s^3 Q_k \omega & 0 &
L(\omega)+ 2J s^4 R_k &
-T_k (\omega)
- J s^4 P_k \\
0 & 2i s^3 Q_k \omega &
-T_{-k}(\omega)
- J s^4 P_{-k} &
L(\omega)+ 2J s^4 R_k
\end{array}
\right)~
\]
 $s$ stands for $\langle S_A^z \rangle $,
 $\omega_0^2 = 4 J^2 s^2 \sin^2(k/2)$.  We
consider for simplicity $J_1=J_2=J$ and the
following definitions are used:
\begin{eqnarray}
Q_k= g_1 g_2 (1 - \cos k) \nonumber \\
R_k=(g_1^2 + g_2^2 - 4g_1g_2)(1+ \cos k) + 2(g_1^2
+ g_2^2) \nonumber \\
P_k =  (1 + \exp [-ik])(g_1 -g_2 \exp [ik])^2 +\nonumber \\
 (1 + \exp [ik])(g_1 - g_2)^2 +  
   4
(g_1 - g_2)(g_1 -g_2 \exp [ik]) \nonumber \\
L(\omega)= (-m \omega^2 +K_1 + K_2)(\omega^2
- \omega_0^2) \nonumber \\
T_k(\omega)= (K_1 + K_2 \exp [ik])
(\omega^2 - \omega_0^2)
\end{eqnarray}

 The eigenvectors of $M$ are given by:

\begin{eqnarray}
{\bf u}^{+} &=& (u_A,u_B;iu_A,iu_B)\label{a_l}\label{u_r}\\
{\bf u}^{-} &=& (u_B^*,u_A^*;-iu_B^*,-iu_A^*)\label{u_l}
\end{eqnarray}
and the amplitudes $u_A$ and $u_B$ are determined by
\begin{eqnarray}
\{ L(\omega) 
-2\omega s^3 Q_k + 
2 J s^4  R_k \} u_A = \nonumber \\
\{ T_k(\omega) +
J s^4 P_k \}u_B 
\end{eqnarray}
  The dispersion relation  $\omega = \omega(k)$
 is given by  the solution of $det M = 0$.
   ${\bf u}^+$ and ${\bf u}^-$ have the same dispersion 
 what translates to the fact that
right and left circular polarized phonons are
degenerate and one might think, that there are
no nonreciprocal effects in antiferromagnets.   
This degeneracy is broken when an external magnetic field is applied 
(see \cite{Boiteux}).
We shall show here that, in fact, even in the absence 
of an external magnetic field, 
nonreciprocal effects might
be obtained by considering an appropriate choice of
boundary conditions.

Let us consider a system that is driven by
a linear polarized amplitude in the $\hat x$-direction
at site $n=0$; {\it i.e.}, we apply the external driving force
\begin{equation}
u_{A,0,x}=\exp[i\omega t]\, u^{(ext.)},\qquad
u_{A,0,y}\equiv 0.
\label{external}
\end{equation}
The appropriate superposition of right and left circular
polarized acoustic waves ${\bf u}^+/{\bf u}^-$
that satisfy (\ref{external}) is
\[
{\bf u}^{}\equiv u_B^*\,{\bf u}^+ +u_A\,{\bf u}^-
=(2u_B^*u_A,|u_B|^2+|u_A|^2;0,i(|u_B|^2-|u_A|^2)),
\]
with $2u_B^*u_A=u^{(ext.)}$. The physical amplitudes,
given by the real part of
$\exp[i(\omega t- k n)]{\bf u}^{}$, are therefore
linearly polarized for the $A$-site,
\[
u_{A,n,x}=\cos(\omega t-kn)\,u^{(ext.)},\qquad
u_{A,n,y}=0, 
\]
and elliptically polarized for the $B$-sites:
\begin{eqnarray}
u_{B,n,x}&=&\cos(\omega t-kn)\,(|u_A|^2+|u_B|^2),
\nonumber\\
u_{B,n,y}&=&\sin(\omega t-kn)\,(|u_A|^2-|u_B|^2).
\nonumber
\end{eqnarray}
The degree of the {\it homogeneous} elliptical
polarization 
$\ (|u_A|^2-|u_B|^2)/(|u_A|^2+|u_B|^2)\sim
  \langle S_A^z\rangle^3 g_1 g_2 (1 - cos k)/(m\omega^3)\ $ 
is proportional to the spin-phonon couplings $g_1$, $g_2$ and
to $\langle S_A^z\rangle$, which makes it nonreciprocal.
Therefore, by choosing appropriate boundary conditions
it is possible to observe this nonreciprocal effect
in an antiferromagnet. Note, in this case, the 
nonreciprocity relates to the amplitudes of the
sound wave and not to the velocity as it happens in the
ordinary acoustic Faraday effect in ferromagnets.

Since the magnetoelastic  coupling between phonons
and spins is determined by general symmetry arguments (see
also \cite{Kittel} and  \cite{Boiteux})
that are not constrained by the dimensionality, we
 can extend our analysis to two and 
three dimensional materials. Then, it follows that the
boundary-condition induced nonreciprocal effect should occur in
materials that consist of antiferromagnetically stacked chains or
planes with each chain/plane having a non vanishing moment along the stacking
direction.  Typical examples are compounds belonging to the 
family of ABX$_3$-type hexagonal antiferromagnets,
like CsCoCl$_3$ and CsNiBr$_3$ \cite{ABX}. The magnitude of
the effect should be comparable to the nonreciprocal
effect observed in ferromagnets \cite{Kittel,Matthews} since 
both effects originate from the same microscopic coupling mechanism.


 \paragraph*{Novel effects}
  In the case illustrated above, we considered a system
 of ions aligned along the $\hat{z}$-axis with cubic
 symmetry in the paramagnetic phase.  Let us now assume
 a crystal with lower symmetry than cubic which also
 orders antiferromagnetically.  We add to our
 Hamiltonian  (\ref{H_sp2}) the following term:
\begin{eqnarray}
H_{sp}^{(f)} =
f_1\sum_n (u_{B,n,x}-u_{A,n,x})
        (S_{B,n}^y S_{A,n}^z+S_{B,n}^z S_{A,n}^y)
\nonumber\\
- f_1 \sum_n (u_{B,n,y}-u_{A,n,y})
   (S_{B,n}^x S_{A,n}^z+S_{B,n}^z S_{A,n}^x)
\nonumber \\
+f_2\sum_n (u_{A,n+1,x}-u_{B,n,x})
        (S_{A,n+1}^y S_{B,n}^z+S_{A,n+1}^z S_{B,n}^y) \nonumber \\
-f_2\sum_n (u_{A,n+1,y}-u_{B,n,y})
        (S_{A,n+1}^x S_{B,n}^z+S_{A,n+1}^z S_{B,n}^x)
\label{H_f}
\end{eqnarray}
where $f_1$ and $f_2$ are, respectively, intra- and intercell
  magnetoelastic coupling constants.  Note that
 this term is invariant under the following symmetries:
(1, $\bar 1$, $2_z$, $\bar 2_z$, $\pm4_z$, $\pm\bar 4_z$) but not,
 for instance, under a reflection $\bar 2_{-xy}$.
 Such a term ($H_{sp}^{(f)}$) can be written for systems crystallizing,
 in the paramagnetic phase, in one of the following point groups:
 $C_{2h}$, $C_{4h}$, $C_{3i}$ or $C_{6h}$ (in Sch\"onflies notation).

  We calculate the
 dispersion relation for $H = T + V_{el} + H_{Heis} + H_{sp}^{(g)}
 + H_{sp}^{(f)}$  by considering the equations of motion for
 $\vec{u}_{A/B,n}$, $\vec{p}_{A/B,n}$
 and $\vec{S}_{A/B,n}$. The staggered magnetization
 axis is taken to be parallel to the $\hat z$-axis
 ($\langle S_{A,n}^z\rangle \rightarrow \langle S_A^z \rangle $
 and
 $\langle S_{B,n}^z\rangle \rightarrow - \langle S_A^z \rangle $).
  The following characteristic polynomial is obtained
\begin{eqnarray}
[m^2 \omega^4 - 2m \omega^2 (K_1 + K_2) + 2 K_1 K_2 (1 - \cos k)]
  [\omega ^2 - \omega_0^2]^2 \nonumber \\
-4m   \langle S_A^z \rangle ^3
 \alpha [\sin k] [\omega ^2 - \omega_0^2 ]\omega ^3 \nonumber \\
- 4  \langle S_A^z \rangle ^6
 [g_1^2 + f_1^2][g_2^2 + f_2^2][1 - \cos k]^2 \omega ^2 \nonumber \\
- 4m  J  \langle S_A^z \rangle ^4
[ (B(3 + \cos k) - 4 \beta (1 + \cos k) ] [\omega^2 - \omega_0^2]
 \omega ^2 \nonumber \\
+ 4 (\omega ^2 - \omega_0 ^2) J \langle S_A^z \rangle ^4
 [3 + \cos k][1 - \cos k]
 [(g_1^2 + f_1^2)K_2 + (g_2^2 + f_2^2)K_1 ] [\omega ^2 - \omega_0 ^2]
  \nonumber \\
+8 J^2 \langle S_A^z \rangle ^8 [g_1^2 + f_1^2] [g_2^2 + f_2^2]
 [1 - \cos k]^3 = 0
\label{novel}
\end{eqnarray}
where $\alpha = g_2 f_1 - f_2 g_1$, $\beta = g_1 g_2 + f_1 f_2$,
 $B = g_1^2 + f_1^2 + g_2^2 + f_2^2 $,   $\omega _0$ is the
 dispersion relation of the unperturbed phonon system and
 $J_1=J_2=J$. While
 the same dispersion relation $\omega = \omega(k)$
  is obtained whether we consider
 right (+) or left (-) circular polarized phonons, Eq.\ (\ref{novel})
 has some important features.  It breaks time reversal symmetry
($\Theta $)  $\omega (k) \neq \omega(-k)$ due to the terms proportional
 to $\alpha \sin k$, and also inversion symmetry (I)
 but respects the product $I \Theta$.
 The breaking
of these two symmetries gives rise to a novel nonreciprocal effect
entirely different from the Faraday effect. It translates to
the fact that transversal
phonon modes propagating in opposite directions have different velocities.
Note that this nonreciprocal effect is directly proportional
to $\alpha$.  If $f_1$ and $f_2$ are zero, there is no such
effect.  

>From a closer inspection of Eq.\ (\ref{H_f}), it
follows that the
three conditions necessary
for the above effect to occur in an antiferromagnet are
(i) the point group should be one of the following:
C$_{2h}$ C$_{4h}$, C$_{3i}$ or C$_{6h}$ \cite{groups},
(ii) the magnetic ions themselves are not centers of
inversion and (iii) the magnetic moment has a component
parallel to the main crystallographic axis, $\langle S^z\rangle\ne0$.
The wolframites \cite{wolframites}
CoWO$_4$, NiWO$_4$, FeWO$_4$ and
FeNbO$_4$ fulfill all three conditions
and are therefore suitable candidates. Their crystal group
is C$_{2h}$ and the magnetic unit cell (2a,b,c) is
doubled with respect to the chemical unit cell.
Another group of candidates are the ilmenites \cite{ilmenites}
FeTiO$_3$, NiTiO$_3$ and MnTiO$_3$ that belong to the
point group C$_{3i}$. 

We estimate for MnTiO$_3$, the relative  difference
\begin{eqnarray}
 \frac{\omega (k) - \omega (-k)}{\omega(k) + \omega (-k)}
\label{difference}
\end{eqnarray}
 for the acoustic phonons. We obtain $\omega(k)$ and $\omega(-k)$ 
 by solving Eq.\ (\ref{novel}) numerically. The
 values of the constants appearing in Eq.\ (\ref{novel})
 are taken from the literature (see \cite{Todate} and \cite{const}).
The magnetoelastic coupling
 constants $g_1, g_2, f_1, f_2$ for manganese ions
 in MnTiO$_3$  are not
 available in the literature. They can be estimated
 from first principles by doing an expansion of the exchange
 term up to the required order \cite{Moriya}. Here,
 for $g_1$  we assume values similar to
 those corresponding to other transition
 metals like Fe.
 We
 assume $f_1$ and $f_2$ to be somewhat smaller than
 $g_1$.  This is reasonable 
given that they define a lower symmetry coupling than $g_1$. 
MnTiO$_3$ has a spin-wave gap $\Delta(H)$ which varies
strongly in an external magnetic field $H$ \cite{Todate}.
We find the novel nonreciprocal effect to be maximal at 
the resonance frequency
of the magnon and the acoustic phonon system which occurs 
at $\omega \simeq 156~\mbox{GHz}$. At resonance the ratio 
Eq.\ (\ref{difference}) is $\sim10^{-4}$. Note that
this nonreciprocal effect occurs already in
absence of an external magnetic field.  In an external
magnetic field the gap $\Delta(H)$ decreases linearly
and vanishes at the spin-flop transition $H_{c}=5.8~\mbox{T}$.
Since the resonance frequency is proportional to the
spin-wave gap, a promising experimental set-up in order to
measure Eq.\ (\ref{difference}) would be
to sweep the external magnetic field for a fixed phonon
frequency. Note that standard measurements of
the velocity of acoustic phonons are accurate to $10^{-5}-10^{-6}$.


\paragraph*{Conclusions}

  We have  studied the possible
 occurrence of an {\it acoustic} Faraday effect in
 antiferromagnets in the absence of
 an external magnetic field by using  a Hamiltonian formulation of the
 magnetoelastic coupling between phonons and spins and
 general symmetry arguments. 
We predict a new boundary- condition induced
homogeneous elliptical polarization effect for layered
materials like the ABX$_3$-type hexagonal antiferromagnets.
  The microscopic analysis of this effect shows that
 it should be of the same order of magnitude as the acoustic Faraday
 effect already observed in ferromagnets \cite{Matthews}.
 We also predict the occurrence of transversal
 phonon modes propagating
 in opposite directions having different velocities
 in certain low-symmetry magnets like the
wolframites or the ilmenites.  An estimate of the order
 of magnitude of this effect for MnTiO$_3$ is given.
 We suggest experiments be undertaken to
detect these effects.

\acknowledgements
We acknowledge useful discussions with A. Auerbach,
B. L\"uthi and G. Shirane.
C.G.\ and R.V.\  thank the hospitality of the Benasque
Center of Physics in Spain. 
R.V.\ acknowledges support through the Deutsche
For\-schungs\-gemein\-schaft. V.N.M.\ thanks the
Graduiertenkolleg {\it Festk\"orperphysik} for
hospitality at Dortmund.
%


%
%
%



\begin{thebibliography}{99}
\bibitem{const}   We take the following estimates: 
$c/3=4.76 \times 10^{-10} m$, $m=2.5 \times 10^{-25} Kg$,
 $J_1 = 1.1214 \times 10^{-22} Joules$,
$J_2 = 1.1025 \times 10^{-22} Joules$,
 $K_1 = 276 Joules/m^2$, $K_2 = 220.8 Joules/m^2$,
 $g_1 = 5 \times 10^6 Joules/m^3$,
 $g_2 = 2.5 \times 10^6 Joules/m^3$, $f_1 = 2.5 \times 10^6 J/m^3$ and
 $f_2 = 2.5 \times 10^6 Joules/m^3$. 
  (all constants are given in units of
 the international system).  The spin of the 
Mn$^{+2}$-ion is $\langle S_A^z \rangle=s=5/2$.
\end{thebibliography}
\end{document}